\documentclass{article}

\usepackage[utf8]{inputenc}
\usepackage[english]{babel}
\usepackage{array}
\usepackage{amsmath,amssymb}
\usepackage{graphics}
\usepackage{graphicx}
\usepackage{color}
\usepackage{lscape}
\usepackage{cleveref}

\def\real{\mathbb{R}}

\def\F{\mathbb{F}}
\def\integer{\mathbb{N}}

\def\A{\mathbb{A}}

\def\L{\mathbb{L}}

\def\mat1{\mathbb{I}}

\def\MEP{Maximum Empower Problem}

\newcommand { \Iletter}[1] {I\kern-0.10em #1 }
\def\bit{\begin{itemize}}
\def\eit{\end{itemize}}
\def\ben{\begin{enumerate}}
\def\een{\end{enumerate}}
\def\bde{\begin{description}}
\def\ede{\end{description}}

\def\bar{\begin{array}}
\def\ear{\end{array}}
\def\beq{\begin{equation}}
\def\eeq{\end{equation}}
\def\bfi{\begin{figure}[hbt] \begin{center}}
\def\efi{\end{center} \end{figure}}
\def\noi{\noindent}
\def\bce{\begin{center}}
\def\ece{\end{center}}
\newcommand {\partie} [1] { \boldsymbol{2}^{#1}}

\newtheorem {theo} {Theorem}

\newtheorem {lem} {Lemma}
\newtheorem{definition}{Definition}[section]

\newcommand{\proof}{{\bf Proof. }}
\newcommand{\cqfd}{\hfill $\Box$ \vskip 0.5cm}

\usepackage[mathcal]{euscript}

\title{A Combinatorial Problem Arising From Ecology: the Maximum Empower Problem}

\author{
Chams Lahlou
\thanks{IMT Atlantique. UBL, Campus de Nantes, La Chantrerie. 4, rue Alfred Kastler - B.P. 20722. F-44307 Nantes Cedex 3. France. (chams.lahlou@imt-atlantique.fr).}
\and Laurent Truffet\thanks{IMT Atlantique. UBL, Campus de Nantes, La Chantrerie. 4, rue Alfred Kastler - B.P. 20722. F-44307 Nantes Cedex 3. France. (laurent.truffet@imt-atlantique.fr).}
}

\begin{document}
\maketitle

\begin{abstract}

The ecologist H. T. Odum introduced a principle of physics, called \textit{Maximum Empower}, in order to explain self-organization in a system (e.g. physical, biological, social, economical, mathematical, ...). The concept of empower relies on \textit{emergy}, which is a second notion introduced by Odum for comparing energy systems on the same basis. The roots of these notions trace back to the 50's (with the work of H. T. Odum and R. C. Pinkerton) and is becoming now an important sustainability indicator in the ecologist community.
In 2012, Le Corre and Truffet developed a recursive method, based on max-plus algebra, to compute emergy of a system. 
Recently, using this max-plus algebra approach, it has been shown that the Maximum Empower Principle can be formalized as a new combinatorial optimization problem (called the Maximum Empower Problem).

In this paper we show that the Maximum Empower Problem can be solved by finding a maximum weighted clique in a cograph, which leads to an exponential-time algorithm in the worst-case.
We also provide a polynomial-time algorithm when there is no cycle in the graph modeling the system.
Finally, we prove that the Maximum Empower Problem is \#P-hard in the general case, i.e. it is as hard as computing the permanent of a matrix.

\end{abstract}


\textbf{AMS:} 90C27, 05C85, 68Q25.

%\linenumbers

%\maketitle

%1) IL Y A DES CAS (DAG) OU LE NOMBRE DE CHEMINS SIMPLES EST EXPONENTIEL MAIS LE CALCUL DU FLOT MAX EST POLYNOMIAL : PEUT-ON AVOIR UN CERTIFICAT DE TAILLE POLYNOMIAL ?
%
%\noindent
%\vskip 0.5cm
%2) VOIR CAS PROBLEME DE DECISION : EXISTE-T-IL UN FLOT DE VALEUR $\geq$ B ?
%
%\noindent
%\vskip 0.5cm
%3) MONTRER LIEN ENTRE CE TYPE DE FLOTS ET LES FLOTS CLASSIQUES ?

\section{Introduction}
The combinatorial optimization problem addressed in this paper is based on a principle of physics which relies on concepts introduced in the mid-50s by the ecologist H. T. Odum and the chemical engineer R. C. Pinkerton \cite{kn:Odum55}: the \textit{Maximum Empower Principle (MEP)}. Since this principle is based on some notions, such as empower and emergy, which are not very known outside the ecologist community, the authors believe that it could be interesting to introduce some basic facts, problems and motivations regarding the MEP.  However, the reader only interested in the mathematical part can go directly to \cref{sub-results}.

\subsection{Some historical notes about emergy, empower and the MEP}
\label{sub-hist}
Scientists have observed since a long time 
that ecological, biological systems, social and economic systems are
energy driven systems (e.g. Podolinsky \cite{kn:Podolinsky1880}; 
Boltzmann \cite{kn:Boltzmann1886}). Nowadays, 
more and more people realize that the world becomes more and more dominated 
by concerns of energy requirements, for sustainable development and 
for environmental conservation. 

Thus, when designing an industrial system, or a human organization, or 
an information system it is important to take into account the 
way energy is used within the system. It means that we have to measure 
the energy efficiency of a system and compare it with another system. In 
other words we need indicators to decide wether or not a system uses 
the energy in an efficient manner. 

The major problem is that complex systems can use energies of different 
kinds. E.g. renewable energies (solar, wind, water,...), 
fossiles energies (fuel, gaz, coal), nuclear energy. Different energies 
are not available at the same time scale. For example, the sun is 
the emergy reference point and is considered to be  
available instantaneously by human being. In this reference system the
 fuel requires several thousands 
of years to be used by human being. Also, different energies do not have the 
same calorific power. 

To address this problem the ecologist Odum proposed the concept of 
emergy (spelled with an 'm'). This term was coined by Scienceman 
in the mid-80's (see e.g. \cite{kn:Scienceman87}). The emergy is defined 
as the ''available energy", also called
 exergy, of one kind used up directly or indirectly to make a 
service or product \cite{kn:Odum96}. Exergy is a thermodynamical 
quantity that 
destruction characterizes the irreversibility of a process (i.e. 
takes into account for energy quality degradation in a process). 
The interested reader is refered to e.g. Moran et al. 
\cite[Chap. 7]{kn:Moranetal}.

To compare systems on the same basis Odum chose as a reference 
the solar emergy, and he introduced 
the notion of transformity. The transformity is defined as the emergy 
required to generate a unit of the available energy in a form different 
from the one of the sun. Thus, 
each type of energy has a transformity. And the reference for 
transformity is the one of the sun, which is equal to one solar 
equivalent Joule per Joule (denoted by 1 sej/J). In other words we 
have:
\begin{equation}
\mbox{Emergy}= \mbox{Transformity} \times \mbox{Exergy},
\end{equation}
where the transformity models the fact that energies are available at different 
time scale and that energies have different calorific power. Note 
that transformities depend on geobiosphere emergy baseline which is 
still a subject of discussion (see e.g. \cite{kn:Brownetal11}, 
\cite{kn:Brownetal2016}, \cite{kn:Campbell2016} and \cite{kn:Vilbissetal2016}).

Because emergy allows us to compare efficiency of different energetic systems on the same 
basis, emergy is becoming an important ecological indicator: e.g. more 
than 1,500 research papers in 10 journals of elsevier. The biggest 
number of contributions mainly comes from USA, China and Italy 
(see e.g. \cite{kn:Chenetal2017}, \cite{kn:Zhongetal2016}). 

Once the reader is convinced of the importance of the emergy in the ecological 
and energetician communities a natural question arises: how to compute 
emergy?  \\

The emergy of a product or a service 
depends on the energy sources of the system and 
how the energy is used by processes within the system to make the product 
or the service. The system is modelled by a graph called in the sequel 
{\em emergy graph}. The emergy graph of a system is a directed weighted 
graph with emergy sources as input nodes, with products, services or consumers 
as output nodes. These nodes fix the boundaries of this 
Multiple Inputs-Multiple Outputs system. It means that the output nodes 
of a system can be the sources of another system. With each source is 
associated an emergy,  which is a positive real number. Its unit is 
the solar equivalent Joule (sej). For example the emergy of the sun during 
one year is estimated to $3.6 \; 10^{24}$ sej (see e.g. 
\cite[p. 18]{kn:Lecorre2016}).

Processes 
within the system are either splits or co-products nodes. A split 
process (or split node) divides input emergy flow as e.g. in hydraulic 
systems. It means that emergy flows after a split are of the same kinds.

A co-product process (or co-product node) divides the input emergy flow 
into emergy flows of different kinds as e.g. in combined heat and power 
plants (see e.g. Horlock \cite{kn:Horlock96}). It means that emergy 
flows after a co-product do not have the same chemical structure.

And the weight of an arc 
between two processes represents the pourcentage of emergy which 
circulates between the two processes. 

Even if in this paper we treat the case of the emergy analysis in 
steady-state assumption, i.e. the emergy of sources 
and the weights of the emergy graph do not depend on time, 
two main difficulties remain:\\

(A) The way the energy is used by 
processes of the interconnected network is modelled by a path which 
represents the 'energy memory' of the product or the 
service from a source. The problem is that only particular paths in 
the graph contribute to the emergy 
of a product or a service. They are called in the sequel {\em emergy 
paths} (see \cref{defEmergyPath}). \\

(B) Because the emergy analysis does not take into account all the paths 
of a graph the Kirchhoff's circuit law at a node of the emergy graph does not
apply.

From (A) and (B) it appears that emergy analysis in steady-state is very
different from e.g. the Leontief input-output approach of economical
models (see e.g. \cite{kn:Leon73}). We retrace hereafter what we think
some important steps and attempts to tackle the emergy analysis
problem. First, note that the emergy evaluation can be divided into
two main steps: (I) the computation of the emergy paths and (II) the
emergy computation as a function of the emergy sources and the emergy
paths.

For step (I). In 1988, Tennenbaum proposed the {\em Track summing method} to 
compute the emergy paths \cite{kn:Tennenbaum88}. This method was only 
described on small examples (see also \cite[Chap. 6]{kn:Odum96}). An attempt to formalize track summing method 
is due to Valyi \cite{kn:Valyi05}. In 2011-2013 Breadth-First and Depth-First Search 
algorithms were used to generate rigourously the emergy paths 
(see Marvuglia et al. \cite{kn:Marvugliaetal2011}, \cite{kn:Scale2013a}).
In 2012, Le Corre and Truffet \cite{kn:Lecorre2012b} proposed an algebraic 
approach to generate the emergy paths based on Benzaken work 
\cite{kn:Benzaken68} (see also e.g. \cite{kn:Carre71}, \cite{kn:BackCarre75}).

For step (II). To the best knowledge of the authors there exist two 
main approaches to compute emergy: (II.a) and (II.b). \\

(II.a). The major attempt 
to enunciate emergy computation rules seems to be due to Brown in 
\cite{kn:Brown96} under the name {\em emergy algebra}. These rules are not 
mathematically formalized (only sentences with sometimes vague terms). 
Let us mention three basic facts concerning emergy algebra: 
\bit
\item ($\alpha$) when splits are reunited the emergy
flows are added (see e.g. \cite{kn:Odum96}, \cite{kn:Lietal2010}, 
\cite{kn:Lazz2009}). 

\item ($\beta$) Co-products, when reunited, cannot be summed. 
Only the emergy of the largest co-product flow is accounted for (see e.g. 
Odum \cite[p.51, Fig. 3.7]{kn:Odum96}). 

\item ($\gamma$) the first 
three rules of emergy algebra do not 
take into account the notion of emergy paths. They are enunciated 
as the Kirchoff's approach of the energy balance of a circuit. 
This fact has generated methods based on linear algebra which provided 
approximate
results. To cite some of them let us mention the Minimum Eigenvalue Model 
\cite{kn:Odum00}, the Linear Optimization Model \cite{kn:Bardietal2005}, and the emergy co-emergy 
analysis \cite{kn:Tennenbaum2014}. These methods did 
not respect all the rules of emergy algebra. In particular, they 
did not treat the co-product problem (see ($\beta$)). Even worse, 
methods based on linear algebra can provide absurd results, i.e. 
negative transformities (see Patterson \cite{kn:Patterson2014}).
That is why some emergeticians introduced methods based on preconditionning 
(see Li et al. \cite{kn:Lietal2010}) or on set theory 
(see Bastianonni et al. \cite{kn:Bast2011}) or Lagrangian approach (see Kazanci et al. \cite{kn:Bastia2012}) or on virtual emergy \cite{kn:Muetal12}. But most 
of these approaches did not provide an automatic treatment of the emergy 
analysis and/or did not respect all the rules of the emergy algebra (specially 
the problem of the co-products).

\eit

In 2012, Le Corre and Truffet reinterpreted the emergy algebra and
proposed a rigourous mathematical framework to compute emergy
\cite{kn:Lecorre2012b}. The emergy in steady-state of a product or a
service is defined as a recursive function $\varphi$ of emergy sources and the
emergy paths. The function $\varphi$ verifies 
six coherent axioms which replace the rules of emergy algebra. And 
because of points ($\alpha$) and ($\beta$) the underlying algebra
is the so called max-plus algebra \cite{kn:Bac-cooq}. The same
authors applied successfuly their method on referenced examples by the
emergy community \cite{kn:Lecorre2012} and on a real world example
\cite{kn:Lecorre2015c}. 

(II.b). Odum proposed the MEP
which claims that: 'In the competition among self-organizing 
processes, network designs that maximize empower will prevail' 
(see e.g. \cite[p. 16]{kn:Odum96},  \cite{kn:Odum95}). Self-organization, 
or spontaneous order principle, states that any
living or non-living disordered system evolves towards an 'equilibrium
state' or coherent state, also called attractor. Self-organization is
observed e.g. in physical, biological, social, mathematical
systems/models, economics, information theory and informatics. The 
empower in steady-state analysis is defined as:

\begin{equation}
\label{eqEmpower}
\mbox{Empower} := \frac{\mbox{Emergy}}{D},
\end{equation}
where $D$ is a given period of time. 

The first attempt to mathematically formalize 
this principle is due to Giannantoni and is based on linear 
algebra and fractional calculus (see e.g. \cite{kn:Giannantoni02}, 
\cite{kn:Giannantoni06}).  
However, as explained in point ($\gamma$), methods based on linear algebra provide approximate results, and sometimes absurd results.

Recently, from the axiomatic basis proposed by Le Corre and Truffet 
\cite{kn:Lecorre2012b}, Lahlou and Truffet \cite{kn:Lahlou2017} 
provided a mathematical formulation of Odum's MEP in steady-state. 
They 
introduced (a) 
the notion of compatible emergy paths of the emergy graph, 
and (b) sets of compatible paths, which they called {\em emergy states}. 
Based on the axiomatic basis of Le Corre and Truffet \cite{kn:Lecorre2012b}, 
they established that: \\
(i) Emergy is mathematically expressed as a maximum over 
all possible emergy states, i.e. it has the following 
form (the precise definition is given in \cref{secPrel}):

\begin{equation}
\label{eqMaxEmp}
\mbox{Emergy} = \max_{\{\hat{\varepsilon}: \mbox{ emergy state}\}}\varphi(\hat{\varepsilon})
\end{equation}
recalling that $\varphi$ is the auxiliary function introduced in 
\cite{kn:Lecorre2012b}. \\
(ii) The maximum is always reached by an emergy state called 
{\em emergy attractor}. \\
(iii) In steady-state, by definition of the empower (see \cref{eqEmpower}), 
the MEP is then restated as follows: 'Only 
prevail emergy states for which the maximum is reached'.

\subsection{Main results and organization of the paper}
\label{sub-results}

In summary, emergy is an ecological indicator developed by the 
ecologist H.T. Odum \cite{kn:Odum96}. It appears to be a way to compare energy driven 
systems efficiency on the same basis, the sun being the reference. 
So, the importance of such ecological indicator is growing up. 

Emergy is a new way to count exergy (defined in e.g. \cite[Chap. 7]{kn:Moranetal}) within a system, and is based on four 
rules called {\em emergy algebra} \cite{kn:Brown96} and a maximization 
principle 
called {\em maximum empower (or emergy) principle} \cite{kn:Odum96}. The way to 
account for emergy is only described by sentences with vague terms. 

In 2012, Le Corre and Truffet \cite{kn:Lecorre2012b} showed that it is possible 
to axiomatize emergy algebra and proposed a recursive definition 
of the emergy in steady-state based on max-plus algebra \cite{kn:Bac-cooq}. 
They applied successfuly their method \cite{kn:Lecorre2012}, 
\cite{kn:Lecorre2015c}. 

In 2017, Lahlou and Truffet \cite{kn:Lahlou2017} showed that the 
recursive definition of the emergy developed in \cite{kn:Lecorre2012b} 
can be seen 
as a maximization problem which could be interpreted as the Odum's 
maximum empower principle. In the sequel, this problem will be 
called the {\em Maximum Empower Problem} (see \cref{subMEP}). 
The conclusion of \cite{kn:Lahlou2017} is that the
formulation of emergy which has the form of (\cref{eqMaxEmp}) appears 
to be a new 
%(to the best knowledge of the authors) 
combinatorial optimization problem on graphs (called emergy graphs, see \cref{defEmergyGraph}) whose complexity has not yet been explored.

\noi
\vskip 0.2cm
{\bf Main results}. 
First, we show that the \MEP\ can be solved by finding a maximum weighted clique in a cograph. This algorithm runs in quadratic time in the number of emergy paths (see \cref{defEmergyPath}), which leads to an exponential-time in the worst-case.
However, we provide a polynomial-time algorithm when the emergy graph is a directed acyclic graph (DAG). Finally, we prove that the \MEP\ is \#P-hard in the general case, i.e. it is as hard as computing the permanent of a matrix. Hence, it seems unlikely to avoid an exponential-time in the worst-case, as it is the case for the proposed algorithm based on a cograph. 
%that this new combinatorial problem is \#P-hard
%in the general case. Roughly speaking, this problem has the same
%complexity as the one of the computation of the permanent of a
%matrix. However, this problem is solvable in polynomial time when . We provide an
%algorithm in this case. We also provide a new algorithm for emergy
%analysis in the general case based on a cograph approach. 

\noi
\vskip 0.2cm
{\bf Organization of the paper}. In \cref{secPrel}, we recall 
some necessary definitions in order to introduce the \MEP.\ 
In \cref{sec:cograph}, we 
present an algorithm for solving the general case. It is based 
on the fact that a cograph can be associated with an emergy graph (see \cref{theo:Cograph}), and on a bijection between weighted cliques in the cograph and solutions of the \MEP\ (see \cref{theo:cograph}). 
In \cref{sec:DAG} we show that when the emergy graph is a DAG 
(possibly with an exponential number of emergy paths) the computation time 
of the emergy is polynomial (see \cref{theo:DAG}). But, in \cref{sec:complexity} we 
show that the Maximum Empower Problem is \#P-hard in the general case 
(see \cref{theo:P-hard}). 
%The reduction is based on a particular  emergy graph which has no co-product nodes. Only split nodes and a 
%general topology (i.e. with cycles). 
Then, in \cref{sec:conclusion} we conclude by some remarks that suggest two open problems.

%%%%%%%%%%%%%%%%%%%%%%%%%%%%%%%%%%%%%%%%%%%%%%%%%%%%%%%%%%%%%%%%%%%%%%%%%%%
\section{Emergy evaluation as a combinatorial optimization problem}
\label{secPrel}

By definition of the empower in steady-state  (see \cref{eqEmpower}) it is 
sufficient to present the main mathematical concepts associated with emergy and
emergy analysis in steady-state. To do this we follow Le Corre and Truffet
\cite{kn:Lecorre2012b}. In the first subsection of this section the notations 
are borrowed from \cite{kn:Lecorre2012b}. %and \cite{kn:Lahlou2017}. 
But to lighten the notations we will present simplified notations 
(see \cref{subNot}). 
Then, in \cref{subMEP}, we formulate the \MEP.

\subsection{Main concepts}
\label{subMC}

An emergy graph $G$ is the following
$10$-tuple:

\begin{equation}
\label{eqEmGraph}
G:=(\L, \L^{s}, \L^{i}, \L^{o}, \mathsf{F}, \A, \mathsf{id}, \perp, \parallel, \emptyset),
\end{equation}
where $\L \subseteq \integer$ denotes the set of all nodes of the 
emergy graph, ($\L^{s}, \L^{i}, \L^{o}$) is a partition of 
$\L$ where $\L^{s}$ denotes the set of emergy 
sources (vertices without predecessors), $\L^{i}$ denotes the set of 
intermediate nodes and $\L^{o}$ denotes the sets of output nodes 
(vertices without successors) which represent services, products 
or consumers. $\mathsf{F}$ is the formal language used to identify 
paths of the graph $G$ with words. Formally, $\mathsf{F}$ is defined as the 
$3$-tuple:
\[
\mathsf{F}:=(\F^{+} \cup \{\underline{0}\} \cup \{\underline{1}\}, \cup, 
\bullet),
\]
where $\F:=\{ [u;v], u,v \in \L\}$, $\F^{+}$ denotes the set of words 
with finite length $\geq 1$ (note that $\F$ is the alphabet), 
$\cup$ denotes the union
operator and $\bullet$ denotes the concatenation of words. $\underline{0}$ 
coincides with the empty set and means that emergy cannot circulate. 
$\underline{1}$ denotes the empty word. $\A \subseteq \F$ denotes the 
set of arcs of the emergy graph $G$. $\mathsf{id}, \perp, \parallel, \emptyset$ 
are symmetric binary relations defined on $\A$ as follows. For $[u;v],[u';v'] 
\in \A$: $[u;v] \mathsf{id} [u';v']$ means $u=u'$ and $v=v'$; 
$[u;v] \emptyset [u';v']$ means that there is no relation between $[u;v]$ 
and $[u';v']$; $[u;v] \parallel [u';v']$ means that there is a co-product 
at $u=u'$; $[u;v] \perp [u';v']$ means that if $u=u'$ 
there is a split at $u$, else $u$ and $u'$ are two different emergy sources.

The emergy evaluation is a path-oriented method thus we introduce or recall hereafter the main definitions concerning emergy paths.

\begin{definition}[Path]
\label{defPath} A path $\pi$ has the form $\pi=\underline{0}$ (if it is the empty path),  or $\pi=\underline{1}$ (if it is a single node), or
$\pi=[l_{1};l_{2}][l_2;l_3] \cdots [l_{k-2};l_{k-1}][l_{k-1};l_{k}]$, with $[l_j;l_{j+1}] \in \A$, for $1\leq j \leq k-1$. 
\end{definition}

\begin{definition}[Concatenation of paths]
\label{defConcatenatePaths} 
The concatenation $\pi \bullet \pi'$ of two paths $\pi$ and $\pi'$ is equal to
\begin{itemize}
\item $\underline{0}$ if $\pi=\underline{0}$ or $\pi'=\underline{0}$.
\item $[l_{1};l_{2}] \cdots [l_{k-1};l_{k}][l'_{1};l'_{2}] \cdots [l'_{k'-1};l'_{k'}]$ if 
$\pi=[l_{1};l_{2}] \cdots [l_{k-1};l_{k}]$, $\pi'=[l'_{1};l'_{2}]$ $\cdots [l'_{k'-1};l'_{k'}]$ and $l_k = l'_1$. 
\end{itemize}
\end{definition}

\begin{definition}[Length of a path]
\label{defLongPath} 
The length of a path $\pi$, $\mathsf{lg}(\pi)$ is the number of arcs which 
compose $\pi$. By convention, $\mathsf{lg}(\underline{0})=- 
\infty $. And $\mathsf{lg}(\underline{1})=0$.
\end{definition}

\begin{definition}[Simple path]
\label{defSimplePath} 
A simple path is a path whose nodes are all different.
\end{definition}

\begin{definition}[Emergy path]
\label{defEmergyPath} 
An emergy path $\pi=[l_{1};l_{2}] \cdots [l_{k-1};l_{k}]$ is a path such 
that $l_{1}$ is an emergy source (i.e. $l_{1} \in \L^{s}$), 
$l_j \in \L \backslash \L^s$, for $2 \leq j \leq k$. And the path 
from $l_{1}$ to $l_{k-1}$ is a simple path.
Notice that the last node $l_k$ may be repeated once.
\end{definition}

Le Corre and Truffet \cite{kn:Lecorre2012b} proposed a recursive definition 
of the emergy $\mathsf{Em}([l;l'])$ flowing on the arc $[l;l']$ of $\A$ 
based on three auxiliary functions: 
\ben 
\item $\theta: \L \rightarrow \real_{+}$ which represents the emergy of a source. We have $\theta(u)>0$ if $u \in \L^{s}$, and $\theta(u)=0$ otherwise.

\item $\omega: \F^{+} \cup \{\underline{0}, \underline{1}\} \rightarrow \real_{+}$. The value $\omega([l_{1};l_{2}])$ represents the pourcentage of emergy flowing between 
processes $l_{1}$ and $l_{2}$ of the emergy graph if $[l_{1};l_{2}] \in \A$. It is equal to $0$ otherwise. 
This function verifies:
\begin{itemize}
\item $\sum_{l_j \in \{\mbox{successors of } l_1\}} \omega_{l_1,l_j} = 1$, if $l_1$ is a split or a source.
\item $\forall l_j \in \{\mbox{successor of } l_1\}$, $\omega_{l_1,l_j} = 1$, if $l_1$ is a co-product node.
\end{itemize}
To learn more about this function we refer the interested reader to e.g. \cite[Section 2]{kn:Lecorre2015c}.
\item $\varphi: \partie{\F^{+}\cup \{\underline{0}, \underline{1}\}} \rightarrow \real_{+}$. This function satisfies 6 axioms (see \cite[Subsection 3.3]{kn:Lecorre2012b}) and defines the emergy flowing on arc $[l;l']$ as:

\begin{equation}
\label{eqEmll'}
\mathsf{Em}([l;l']) := \varphi(\varepsilon([l;l'])),
\end{equation} 
where
$\varepsilon([l;l'])$ denotes the set of all emergy paths of the 
emergy graph $G$ ending by the arc $[l;l']$.

. 
%but in this paper only two of them will be used (see  \cite[Subsection 3.3]{kn:Lecorre2012b}):
%\begin{description}
%\item[-] For a path $\pi$ and a set of paths $\U$:
%\begin{equation}
%\label{eq:varphi0a}
%\varphi(\{\pi \bullet \pi' : \pi' \in \U\}) = \varphi(\pi) \varphi(\{\pi' : \pi' \in \U)\}). 
%\end{equation}
%\item[-] Let $l_1$ be a split node, and let $[l_1;l_j]$ be some arcs of $\A$, for $1 \leq j \leq k$. If $\U_j$ is a set of paths starting by node $l_j$ and ending by $[l;l']$ then
%\begin{equation}
%\label{eq:varphi0b}
%\varphi(\bigcup_{1 \leq j \leq k} \{[l_1;l_j] \bullet \pi : \pi \in \U_j )\}) = \sum_{1 \leq j \leq k} \varphi(\{[l_1;l_j] \bullet \pi : \pi \in \U_j )\}).
%\end{equation}
%\end{description}

\een

\subsection{Notations}
\label{subNot}

In the rest of the paper it will be convenient to use graph-oriented notations rather than the formal language-oriented notation used in \cite{kn:Lecorre2012b}:

\bit
\item An arc $[l_{1};l_{2}]$ will be simply denoted by $(l_{1},l_{2})$.
% or $l_{1}l_{2}$.

\item A path $\pi=[l_{1};l_{2}] \cdots [l_{k-1};l_{k}]$ will be denoted by $(l_{1}, \ldots, l_{k})$. 
The concatenation $\pi \bullet \pi'$ of two paths $\pi$ and $\pi'$ will be denoted by $\pi\pi'$.
%(i.e. if $\pi=(l_{1}, \ldots, l_{k})$ and $\pi'=(l'_{1}, \ldots, l'_{k'}), to a path$\pi$ will be denoted $\pi\pi'$

\item The value $\omega([l_{1};l_{2}])$ of an arc $(l_{1},l_{2})$ will be denoted 
by $\omega_{l_{1},l_{2}}$.

\item The emergy $\mathsf{Em}([l;l'])$ flowing on arc $(l,l')$ will be denoted by $\mathsf{Em}(l,l')$.

\item The set  $\varepsilon([l;l'])$ of emergy paths ending by arc $(l,l')$
will be denoted by $\varepsilon(l,l')$. 

\item The set $\Gamma^+(i)$ will denote the successors of a vertex $i$.
\eit 

Moreover, we can remark that the set $\L$ of the nodes of $G$ can be also partitioned into four sets of nodes defined hereafter.

\begin{figure}[htbp]
\centering
\includegraphics[scale=0.7]{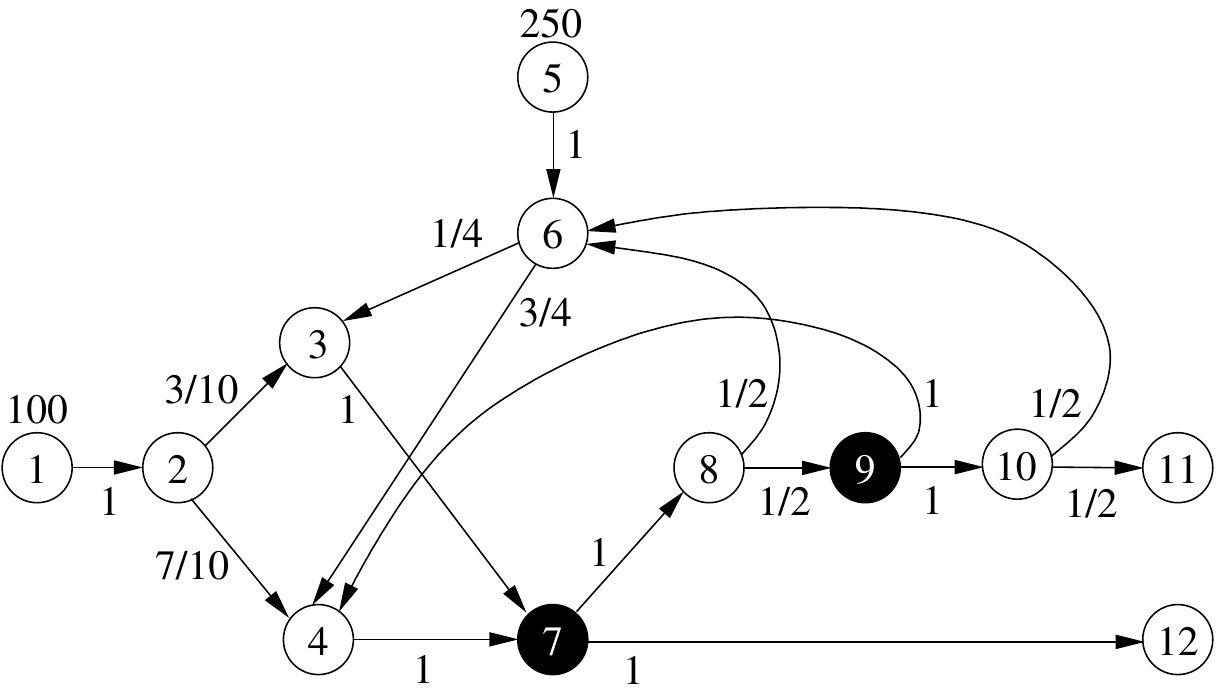}
\caption{An emergy graph (after example of \cite[p. 27]{kn:Lecorre2016}). Co-product nodes are in black. The number placed above a source node $s$ represents its weight $\theta(s)$; The number placed along an arc $(i,j)$ represents its weight $\omega_{i,j}$.}
\label{fig:example}
\end{figure}

\begin{definition}[Emergy source node]
It is an element of $\L^{s}$. It means that it is a vertex without predecessors. 
By convention, each source $s$ of the emergy graph is connected to only one node of the emergy graph denoted by $\mathsf{succ}(s)$. 
\end{definition}
\begin{definition}[Split node]
It is an element $u$ of $\L^{i}$ which has 
$u_{1}, \ldots, u_{k}$ successors for some $k \geq 1$; when $k \ge 2$, it is such that $(u,u_j) \perp (u,u_{j'})$ for $1 \leq j < j' \leq k$ . 
\end{definition}
\begin{definition}[Co-product node]
It is an element $u$  of $\L^{i}$ which has 
$u_{1}, \ldots, $ $u_{k}$ successors, for some $k \geq 2$, and such that $(u,u_j) \parallel (u,u_{j'})$ for $1 \leq j < j' \leq k$. 
\end{definition}
\begin{definition}[Output node]
It is an element of $\L^{o}$. It means 
that it is a vertex without successors. 
%It is an output node of the system.
\end{definition}
For the example given in \cref{fig:example} we have $V_s=\{1,5\}$, $V_+=\{2,3,4,6,8,10\}$, $V_{\max}=\{7,9\}$ and $V_o=\{11,12\}$.
%%%%%%%%%%%%%%%%%%%%% FIGURE EXEMPLE %%%%%%%%%%

%We can also consider that $\theta(s)$ is the weight associated with the source node $s$, and $\omega_{i,j}$ is the weight of arc $(i,j)$.

%Finally, denoting these sets by $V_s$,$V_+$,$V_{\max}$ and $V_o$ respectively, definition \eqref{eqEmGraph} can be simplified into  
%\begin{equation}
%\label{eqEmGraph2}
%G:=(V_s, V_+, V_{\max}, V_o,A)
%\end{equation}
%where $A$ is the set of arcs $\A$.

Finally, we use this partition to introduce a new definition of an emergy graph which includes functions $\theta$ and $\omega$ as weights.
\begin{definition}[Emergy graph]
\label{defEmergyGraph}
It is a weighted directed graph $(V_s, V_+, V_{\max},$ $V_o, A, \theta, \omega)$ where: 
$V_s$,$V_+$,$V_{\max}$ and $V_o$ are the sets of source nodes, split nodes, co-products nodes and output nodes, respectively;
$A$ is the set of arcs;
$\theta(s)$ is the (nonnegative) weight of source node $s$;
$\omega_{i,j}$ si the (nonnegative) weight of arc $(i,j)$;
\end{definition}

\subsection{The \MEP}
\label{subMEP}

Lahlou and Truffet \cite{kn:Lahlou2017} proved that the recursive definition of the emergy given in \cite{kn:Lecorre2012b} (and expressed by \cref{eqEmll'}) is equivalent to a combinatorial optimization problem called the {\em \MEP}.
To establish this result, they showed that only some paths are taken into account for computing the emergy flowing on an arc. For that, they introduced the notions of compatible paths and emergy state.

\begin{definition}[Compatible paths]
\label{def:CompatiblePaths}
Let $\pi=(l_{1},l_{2}, \ldots, l,l')$ and $\pi'=(l_{1}',l_{2}',$ $
\ldots,l,l')$ be two paths ending by an arc $(l,l')$. They are compatible relatively to the arc $(l,l')$, which is denoted by $\pi \hat{+} \pi'$, if and only if one of the following cases occurs:\\
(i). $\pi = \pi'$.\\
(ii). $l_1$ and $l'_1$ are two different sources.\\
(iii). $\exists k$, with $k \ge 1$, such that $l_{i}= l_{i}'$, for $1 \leq i \leq k$,  and $l_{k}$ is a split node.
\end{definition}

For the example given in \cref{fig:example}, paths $(3,7,8,6,4)$ and $(3,7,8,9,10,6,4)$ are compatible (case (iii) applies with $k=3$), whereas paths $(8,9,4,7)$ and $(8,9,10,6,4,7)$ are not compatible because they divide at node 9 which is a co-product.

\begin{definition}[Emergy state]
An emergy state $\hat{\varepsilon}$, relatively to an arc $(l,l')$,  is a set of pairwise compatible emergy paths relatively to $(l,l')$.
\end{definition}
%Let $i$ be a split node and let $\mathbb{U}_j$ be a set of paths starting from node $j$, where $j$ is a successor of $i$. Function $\varphi$ applied to $\cup_{j \in \Gamma^+(i)} \{(i,j)\pi : \pi \in \U_j )\}$ verifies (see \cite{kn:Lecorre2012b})
%\begin{equation}
%\label{eq:varphi0a}
%\varphi(\cup_{j \in \Gamma^+(i)} \{(i,j)\pi : \pi \in \U_j )\}) = \sum_{j \in \Gamma^+(i)} \varphi(\{(i,j)\pi : \pi \in \U_j \}), 
%\end{equation}
%and, for a set of paths $\U$,
%\begin{equation}
%\label{eq:varphi0b}
%\varphi(\{\pi \pi': \pi' \in \U \}) = \varphi(\pi)  \varphi(\{\pi': \pi' \in \U \}). 
%\end{equation}
Using these definitions, they proved that function $\varphi$ applied to an an emergy state $\hat{\varepsilon}$ verifies (see \cite[Corollary 1 and Proposition 4]{kn:Lahlou2017})
%\begin{subequations}
\begin{equation}
\label{eq:varphi1}
\varphi(\hat{\varepsilon}) = \sum_{\pi \in \hat{\varepsilon}} \varphi(\pi), 
\end{equation}
where, for a path $\pi$ (see \cite[Subsection 3.3]{kn:Lecorre2012b}),
\begin{equation}
\label{eq:varphi2}
\varphi(\pi) = 
\left\{
\begin{array}{ll}
0 &  \mbox{if } \mathsf{lg}(\pi)=-\infty \mbox{ (i.e. } \pi=\underline{0}), \\
1 & \mbox{if } \mathsf{lg}(\pi)=0 \mbox{ (i.e. } \pi=\underline{1}), \\
\theta(s) \prod_{(i,j) \in \pi} \omega_{i,j} & \mbox{if } \mathsf{lg}(\pi) \ge 1 \mbox{ and } \pi\mbox{ starts by a source node } s, \\
\prod_{(i,j) \in \pi} \omega_{i,j} & \mbox{if } \mathsf{lg}(\pi) \ge 1 \mbox{ and } \pi \mbox{ does not start by a source}.\\
\end{array}
\right.
\end{equation}
%\end{subequations}

%\begin{rem}
%\label{rem:wij}
%The function $\omega$ is a particular nonnegative function. It verifies 
%that $\sum_{j \in \Gamma^+(i)} \omega_{i,j} = 1$ if $i$ is a split node, and $\forall j \in \Gamma^+(i), \omega_{i,j} = 1$, if $i$ is a co-product node. To learn 
%more about this function we refer the interested 
%reader to e.g. \cite[Section 2]{kn:Lecorre2015c}.
%\end{rem}

Then, they proved that 
\begin{equation}
\label{eq:Optim}
\mathsf{Em}(l,l')= \max_{\hat{\varepsilon} \in \hat{E}(l,l')}\varphi(\hat{\varepsilon}),
\end{equation}
where $\hat{E}(l,l')$ (with $\hat{E}(l,l') \subseteq \partie{\varepsilon(l,l')}$) denotes the set of all emergy states relatively to the arc $(l,l')$. 

This leads to the following definition of the optimization problem:
\begin{definition}[\MEP]
Given an emergy graph $(V_s, $ $V_+, V_{\max}, V_o, A,\theta,\omega)$, an arc $(l,l')$ of $A$, and function $\varphi$ defined by \cref{eq:varphi2} solve
\begin{equation}
\label{eq:MEP}
\max_{\hat{\varepsilon} \in \hat{E}(l,l')} \sum_{\pi \in \hat{\varepsilon}} \varphi(\pi)
\end{equation}
\end{definition}
\noindent
\textbf{Example.}\\
Let us go back to the example given in \cref{fig:example}, and let us solve the associated \MEP\ when $(l,l')$ is set to $(4,7)$.
There are 4 emergy paths from source 1: $(1,2,4,7)$, $(1,2,3,7,8,6,4,7)$, $(1,2,3,7,8,9,4,7)$ and $(1,2,3,7,8,9,10,6,4,$ $7)$. Notice that the first path is compatible with the other three paths.
And there are 2 emergy paths from source 5 ($(5,6,4,7)$ and $(5,6,3,7,8,9,4,7)$) which are also compatible.

Since two emergy paths starting from different sources are compatible (case (ii) of \cref{def:CompatiblePaths}), and since $\varphi$ is a nonnegative function, an optimum emergy state $\hat{\varepsilon}^*$ must contain the first path starting from source 1 and both paths starting from source 5, i.e. $\{(1,2,4,7), (5,6,4,7), (5,6,3,7,8,9,4,7)\} \subseteq \hat{\varepsilon}^*$. Since the three remaining paths starting from source 1 are not compatible, but each of them is compatible with every path of $\{(1,2,4,7), (5,6,4,7), (5,6,3,7,8,9,4,7)\}$, only one among them can belong to $\hat{\varepsilon}^*$. So let us compute the value of $\varphi$ for them: \\
- $\varphi((1,2,3,7,8,6,4,7))= 100 \cdot 1 \cdot \frac{3}{10} \cdot 1 \cdot 1 \cdot \frac{1}{2} \cdot \frac{3}{4} \cdot 1 = \frac{45}{4}$,\\
- $\varphi((1,2,3,7,8,9,4,7))= 100 \cdot 1 \cdot \frac{3}{10} \cdot 1 \cdot 1 \cdot \frac{1}{2} \cdot 1 \cdot 1 = 15$,\\
- $\varphi((1,2,3,7,8,9,10,6,4,7))= 100 \cdot 1 \cdot \frac{3}{10} \cdot 1 \cdot 1 \cdot \frac{1}{2} \cdot 1 \cdot  \frac{1}{2} \cdot \frac{3}{4} \cdot 1 = \frac{45}{8}$.\\
Hence, we deduce that $\hat{\varepsilon}^*= \{(1,2,4,7), (5,6,4,7), (5,6,3,7,8,9,4,7), (1,2,3,7,$ $8,9,4,7)\}$, and we get $\varphi(\hat{\varepsilon}^*) = (100 \cdot 1 \cdot \frac{7}{10} \cdot 1)+ (250 \cdot 1 \cdot \frac{3}{4} \cdot 1) + (250 \cdot 1 \cdot \frac{1}{4} \cdot 1 \cdot 1 \cdot \frac{1}{2} \cdot 1 \cdot 1 ) + 15 = 70 + \frac{375}{2} + \frac{125}{4} + 15 = 303.75$ sej.

%%%%%%%%%%%%%%%%%%%%%%%%%%%%%%%%%%%%%%%%%%%%%%

\section{The maximum empower problem as a maximum weighted clique in a cograph}
\label{sec:cograph}

%Example (Figure \ref{fig:network}): we have
%$P_4=\{(4,6,9,t),(4,6,2,3,5,9,t)\}$. The longest common prefix of
%$P_4$ is the path $(4,6)$.

%%%%%%%%%%%%%%%%%%%%%%%%%%%%%%%%%%%%%%%%%%%%%%

There are numerous definitions of a cograph (see e.g. \cite{Corneil-Lerchs-StewartBurlingham1981}). We use the following one:

\begin{definition}
A cograph is a graph that does not contain a path of four vertices as an induced subgraph.
\end{definition}
Hence, if a cograph contains a path of the form $(a,b,c,d)$ it must contain at least one of the edges $\{a,c\}$,$\{a,d\}$ or $\{b,d\}$.
We shall use this property to show that a cograph can be associated with an emergy graph. Our proof makes also use of the following definition:

\begin{definition}
Let $P$ be a set of emergy paths $\pi_1, \ldots,\pi_k$ that start from the same vertex, say $i$, and end by the same arc.
The longest common prefix of $P$ is the longest path $\pi$ that starts from $i$, and such that there exist $\pi'_1, \ldots,\pi'_k$ with $\pi_j=\pi \pi'_j$ for $1\leq j \leq k$.
\end{definition}

Notice that this definition together with \cref{def:CompatiblePaths} imply that two emergy paths (ending by the same arc) are not compatible if and only if they start from the same source and the last vertex of their longest common prefix is a co-product.
%%%%%%%%%%%%%%%%%%%%% FIGURE COGRAPH %%%%%%%%%%
\begin{figure}[htbp]
\centering
\includegraphics[scale=0.8]{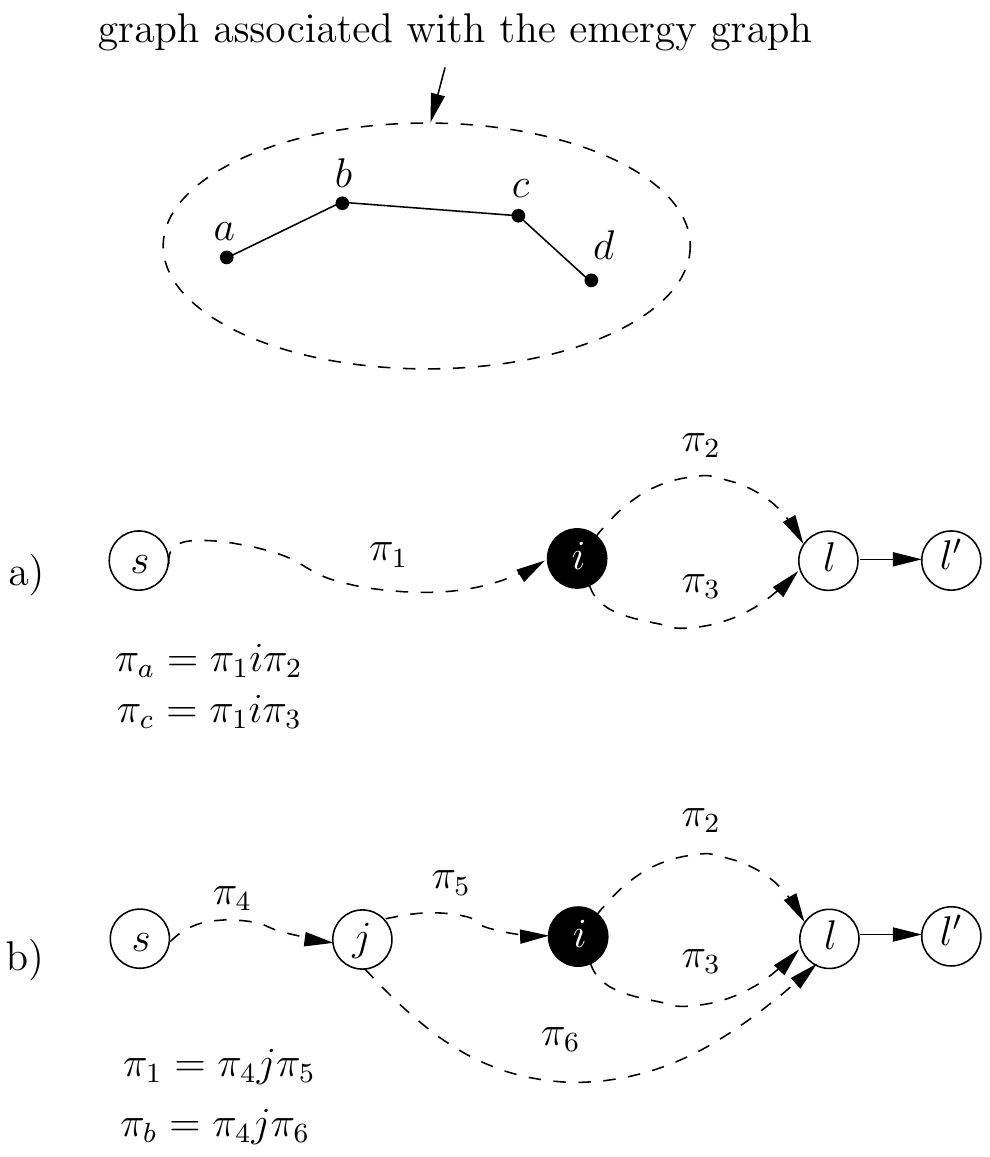}
\caption{The argument used for showing that the set of emergy paths defines a cograph.}
\label{fig:cograph}
\end{figure}

\begin{theo}
\label{theo:Cograph}
An emergy graph and a given arc define a cograph as follows: a vertex is associated with each emergy path, and an edge is associated with each pair of vertices if they represent emergy paths that are compatible relatively to the arc.
\end{theo}

\proof
Assume, for the sake of contradiction, that the associated graph
contains a path $(a,b,c,d)$ as an induced subgraph.  Let $\pi_a, \pi_b, \pi_c$
and $\pi_d$ be the associated emergy paths.

Since $a$ and $c$ are not joined by an edge, paths $\pi_a$ and $\pi_b$ are
not compatible, that is they start from the same source and the last vertex of their longest common prefix is a co-product. Let $i$ be this vertex. Consider a decomposition into
subpaths such that $\pi_a=\pi_1i \pi_2$ and $\pi_c=\pi_1i \pi_3$ (see \cref{fig:cograph}.a).

Since $b$ is connected to $a$ and $c$, path $\pi_b$ is compatible with
both $\pi_a$ and $\pi_c$. Let $j$ and $j'$ be the last vertex of the
longest common prefix of $\pi_b$ and $\pi_a$, and of $\pi_b$ and $\pi_c$,
respectively. Since vertices $j$ and $j'$ are split vertices, we have
$j \neq i$ and $j' \neq i$.  If $j$ belongs to $\pi_2$ then paths $\pi_b$
and $\pi_c$ separate at vertex $i$, i.e. $j'=i$, which is a
contradiction. Hence, $j$ belongs to $\pi_1$.  Similarly, if $j'$
belongs to $\pi_3$, we have $j=i$, which is a contradiction. Thus, $j'$
belongs to $\pi_1$ and we have $j=j'$. Let us decompose $\pi_1$ 
into $\pi_4 j \pi_5$, and $\pi_b$ into $\pi_4j\pi_6$ (see \cref{fig:cograph}.b).

Now, let $k$ be the last vertex of the longest common prefix of $\pi_b$
and $\pi_d$. Since it is a co-product ($b$ and $d$ are not connected) we
must have $k \neq j$, which implies that $k$ belongs to $\pi_4$ or $\pi_6$. 
In the former case, $k$ is also the last vertex of the longest common prefix of $\pi_d$ and $\pi_c$, which leads to a contradiction since $k$ must then be a split vertex ($d$ and $c$ are connected). In the latter case, we have again a contradiction since the last vertex of the longest common prefix of $\pi_d$ and $\pi_a$ is then vertex $j$, which implies that $d$ and $a$ are connected.
Hence, vertex $k$ cannot exist, which concludes the proof.~\cqfd

Now, we are able to propose an algorithm for our problem.

\begin{theo}
\label{theo:cograph}
The \MEP\ is solved by the following algorithm:

\begin{enumerate}
\item Compute the set of emergy paths $\varepsilon(l,l')$.
\item Create a graph $G$ whose sets of vertices is $\varepsilon(l,l')$ and whose set of edges is $\{\{\pi,\pi'\} : \pi \in \varepsilon(l,l'), \pi' \in \varepsilon(l,l'), \pi \hat{+} \pi' \}$.
\item Associate with each vertex $\pi$ of $G$ a weight equal to $\varphi(\pi)$.
\item Compute a maximum weighted clique of $G$.
\end{enumerate}

\end{theo}

\proof
By construction there is a bijection between cliques of $G$ and emergy
states. Moreover, the weight of a clique $K$ is equal to $\sum_{\pi \in K} \varphi(\pi)$, which is the value of the associated emergy state (recall Definitions \eqref{eq:varphi1} and \eqref{eq:varphi2}).
Since graph $G$ is a cograph (by \cref{theo:Cograph}) and since Jung \cite{Jung1978} proved that a cograph is also a comparability graph, we can use the algorithm of Golumbic \cite[p. 314]{Berge-Chvatal1984} for finding a maximum weighted clique of a comparability graph.~\cqfd

Because Golumbic's algorithm runs in linear time (with respect to the number of vertices and edges of the graph) and since it takes time ${\cal O}(|V_s \cup V_+ \cup V_{\max} \cup V_o|)$ to check if two paths are compatible, the algorithm has a time complexity of ${\cal O}(|\varepsilon(l,l')|^2)$.
Obviously, it means an exponential time in the worst case. 
However, we shall see in the following section that when there are no cycles, that is when the emergy graph is a directed acyclic graph (DAG), it is possible to solve the \MEP\ in polynomial time, even if there is an exponential number of emergy paths.

%%%%%%%%%%%%%%%%%%%%%%%%%%%%%%%%%%%%%%%%%%%%%%
\section{Computing the maximum empower in polyno-\\ mial-time when the emergy graph is a DAG}
\label{sec:DAG}
Let $G$ be an emergy graph defined by $(V_s, V_+, V_{\max}, V_o, A,$ $ \theta, \omega)$, and let $(l,l') \in A$. When $G$ is a DAG a path from a source to the arc $(l,l')$ is always a simple path, which means that it is also an emergy path. Our approach is based on this remark. 

Let $\varepsilon_i$ denote the set of simple paths starting from $i$ and ending by $(l,l')$, and let $\hat{E}_i=\{\hat{e} :  \hat{e} \subseteq \varepsilon_i \mbox{ and } \forall \pi \in \hat{e}, \forall \pi' \in \hat{e}, \pi \hat{+} \pi' \}$.

The algorithm is based on the function $f$ defined for every node $i$ of $G$ as follows:

\begin{itemize}
\item If $i$ is a node such that $\varepsilon_i = \emptyset$ then
\begin{equation}
\label{eq:DAGfunction0}
f(i) = 0,
\end{equation}
\item else
\begin{equation}
\label{eq:DAGfunction}
f(i) = 
\left\{
\begin{array}{ll}
\theta(i) & \mbox{ if }  i= l \mbox{ and } i \in V_s, \\
\omega_{i,l'} & \mbox{ if } i= l \mbox{ and } i \notin V_s, \\
\theta(i)f(\mathsf{succ}(i)) & \mbox{ if } i \neq l \mbox{ and } i \in V_s, \\
\sum_{j \in \Gamma^+(i)} \omega_{i,j} f(j) & \mbox{ if } i \neq l \mbox{ and } i \in V_+, \\
\max_{j \in \Gamma^+(i)} f(j) & \mbox{ if }  i \neq l \mbox{ and }  i \in V_{\max}. \\
\end{array}
\right.
\end{equation}
\end{itemize}

It solves a particular maximization problem:
\begin{lem}
\label{lem:DAG}
Given an arc $(l,l')$ of the emergy graph and a vertex $i$ we have
\begin{equation}
\label{eq:DAG}
\max_{\hat{e} \in \hat{E}_i} \sum_{\pi \in \hat{e}} \varphi(\pi) = f(i)
\end{equation}
\end{lem}

\proof 
If $\varepsilon_i = \emptyset$, then no path that starts from $i$ can end by arc $(l,l')$. Hence, $\hat{E}_i=\emptyset$, so $\max_{\hat{e} \in \hat{E}_i} \sum_{\pi \in \hat{e}} \varphi(\pi) = \varphi(\underline{0})$, i.e. $\max_{\hat{e} \in \hat{E}_i} \sum_{\pi \in \hat{e}} \varphi(\pi) = 0$ by \eqref{eq:varphi2}, and relation \eqref{eq:DAG} is true. 

So assume that $\varepsilon_i \neq \emptyset$ and let us prove \eqref{eq:DAG} by induction on the length $\mathsf{lg}(\pi)$ of a longest path $\pi$ of $\varepsilon_i$. 
Since $\varepsilon_i \neq \emptyset$, any path of $\varepsilon_i$ contains at least arc $(l,l')$, which means that $\max_{\pi \in \varepsilon_i} \mathsf{lg}(\pi) \ge 1$ .

If $\max_{\pi \in \varepsilon_i} \mathsf{lg}(\pi)=1$, then $i=l$ and $\hat{E}_i=\{(i,l')\}$, so $\max_{\hat{e} \in \hat{E}_i} \sum_{\pi \in \hat{e}} \varphi(\pi) = \varphi((i,l'))$. 
If $i \in V_s$, $\max_{\hat{e} \in \hat{E}_i} \sum_{\pi \in \hat{e}} \varphi(\pi) = \theta(i) \omega_{i,l'}$ by Definition \eqref{eq:varphi2} of $\varphi$, i.e. $\max_{\hat{e} \in \hat{E}_i} \sum_{\pi \in \hat{e}} \varphi(\pi) = \theta(i)$ since $i$ is a source (recall definition of $\omega$).
Else, $\max_{\hat{e} \in \hat{E}_i}$ $\sum_{\pi \in \hat{e}} \varphi(\pi) = \omega_{i,l'}$ by Definition \eqref{eq:varphi2} of $\varphi$.
Thus, \eqref{eq:DAG} is true.

If $\max_{\pi \in \varepsilon_i} \mathsf{lg}(\pi) \ge 2$ then $i \neq l$.
Assume that for any $i$ such that $\max_{\pi \in \varepsilon_i} \mathsf{lg}(\pi)$ $ \leq k$, with $k \ge 1$, we have 
$f(i) = \max_{\hat{e} \in \hat{E}_i}$ $\sum_{\pi \in \hat{e}} \varphi(\pi)$. Let us consider a vertex $i$ such that  $\max_{\pi \in \varepsilon_i} \mathsf{lg}(\pi) = k+1$.
We have three cases:

\begin{description}
\item[Case 1:] $i \in V_s$.
Since a source has only one successor we have $\hat{e} \in \hat{E}_i$ if and only if there exists $\hat{e}'$ such that $\hat{e}' \in \hat{E}_{\mathsf{succ}(i)}$ and $\hat{e} = \{(i,\mathsf{succ}(i)) \pi' : \pi' \in \hat{e}' \}$. Thus, we have
$$
\max_{\hat{e} \in \hat{E}_i} \sum_{\pi \in \hat{e}} \varphi(\pi) = \max_{\hat{e}' \in \hat{E}_{\mathsf{succ}(i)}} \sum_{\pi' \in \hat{e}'} \varphi((i,\mathsf{succ}(i))\pi'),
$$
i.e., by Definition \eqref{eq:varphi2} of $\varphi$,
$$
\max_{\hat{e} \in \hat{E}_i} \sum_{\pi \in \hat{e}} \varphi(\pi) = \max_{\hat{e}' \in \hat{E}_{\mathsf{succ}(i)}} \sum_{\pi' \in \hat{e}'} \varphi((i,\mathsf{succ}(i))) \varphi(\pi').
$$
Since $i \in V_s$, we have $\varphi((i,\mathsf{succ}(i))) = \theta(i)$ (by Definition \eqref{eq:varphi2} of $\varphi$) and, since $\theta(i) \ge 0$, we get
$$
\max_{\hat{e} \in \hat{E}_i} \sum_{\pi \in \hat{e}} \varphi(\pi) = \theta(i) \max_{\hat{e}' \in \hat{E}_{\mathsf{succ}(i)}} \sum_{\pi' \in \hat{e}'} \varphi(\pi').
$$
Because $\mathsf{lg}(\pi') \leq k$, for $\pi' \in \hat{e}'$, the hypothesis of induction applies, so
$$
\max_{\hat{e} \in \hat{E}_i} \sum_{\pi \in \hat{e}} \varphi(\pi) = \theta(i)f(\mathsf{succ}(i)).
$$
We conclude that \eqref{eq:DAG} is true.

\item[Case 2:]  $i \in V_+$.
If $i$ has only one successor the reasoning of Case 1 applies and \eqref{eq:DAG} is true.
So let us assume that $i$ has more than one successor and assume, without loss of generality, that $\Gamma^+(i)=\{1, \ldots,\gamma\}$ where $\gamma=|\Gamma^+(i)|$.

Let $j$ and $j'$ be two successors of $i$, and let $\hat{e} \in \hat{E}_j$ and $\hat{e}' \in \hat{E}_{j'}$. 
Since $i$ is a split, every path of $\{(i,j)\pi : \pi \in \hat{e}  \}$ is compatible with every path of $\{(i,j')\pi' : \pi' \in \hat{e}'  \}$.
Therefore, $\hat{e} \in \hat{E}_i$ if and only if there exist $\hat{e}_j \in \hat{E}_{j} \cup \{\emptyset\}$, for $1\leq j \leq \gamma$, such that  $\hat{e} = \cup_{j=1}^{\gamma} \{(i,j) \pi_j : \pi_j \in \hat{e}_j\}$. 
Recalling that $\emptyset \notin \hat{E}_{i}$ (since $\varepsilon_i \neq \emptyset$) and setting $E=(\hat{E}_{1} \cup \{\emptyset\}) \times \cdots \times (\hat{E}_{\gamma} \cup \{\emptyset\}) \backslash (\emptyset, \ldots, \emptyset)$, we have
%Let $\{ (\hat{e}_j) : \hat{e}_j \in \hat{E}_j, j \in \Gamma^+(i) \}$ be the set of tuples of this bijection. 
$$
\max_{\hat{e} \in \hat{E}_i} \sum_{\pi \in \hat{e}} \varphi(\pi)
=
\max_{(\hat{e}_1, \ldots,\hat{e}_{\gamma}) \in E} \sum_{\pi \in \cup_{j=1}^{\gamma} \{(i,j)\pi_j: \pi_j \in \hat{e}_j \}}  \varphi(\pi) .
$$
Since $\hat{e}_j \cap \hat{e}_{j'} = \emptyset$, for $1 \leq j < j' \leq \gamma$, we get
$$
\max_{\hat{e} \in \hat{E}_i} \sum_{\pi \in \hat{e}} \varphi(\pi)
=
\max_{(\hat{e}_1, \ldots,\hat{e}_{\gamma}) \in E} \sum_{1\leq j \leq \gamma} \ \sum_{\pi_j \in \hat{e}_j}  \varphi((i,j)\pi_j),
$$
i.e., by Definition \eqref{eq:varphi2} of $\varphi$,
$$
\max_{\hat{e} \in \hat{E}_i} \sum_{\pi \in \hat{e}} \varphi(\pi)
=
\max_{(\hat{e}_1, \ldots,\hat{e}_{\gamma}) \in E} \sum_{1\leq j \leq \gamma} \ \omega_{i,j}\sum_{\pi_j \in \hat{e}_j}  \varphi(\pi_j).
$$
Applying the distributivity of operator + over $\max$ we get
$$
\max_{\hat{e} \in \hat{E}_i} \sum_{\pi \in \hat{e}} \varphi(\pi)
=
\sum_{1\leq j \leq \gamma} \max_{\hat{e}_j \in \hat{E}_j \cup \emptyset}  \omega_{i,j}\sum_{\pi_j \in \hat{e}_j}  \varphi(\pi_j).
$$
Since $\omega_{i,j} \ge 0$, for $(i,j) \in A$, we obtain
$$
\max_{\hat{e} \in \hat{E}_i} \sum_{\pi \in \hat{e}} \varphi(\pi)
=
\sum_{1\leq j \leq \gamma} \omega_{i,j} \max_{\hat{e}_j \in \hat{E}_j \cup \emptyset}  \sum_{\pi_j \in \hat{e}_j}  \varphi(\pi_j).
$$
Finally, since $\varphi(\{\emptyset\})=0$ and $\varphi \ge 0$, it is equivalent to
$$
\max_{\hat{e} \in \hat{E}_i} \sum_{\pi \in \hat{e}} \varphi(\pi)
=
\sum_{1\leq j \leq \gamma} \omega_{i,j} \max_{\hat{e}_j \in \hat{E}_j}  \sum_{\pi_j \in \hat{e}_j}  \varphi(\pi_j).
$$
Then, by the hypothesis of induction (which applies since $\mathsf{lg}(\pi_j) \leq k$ for $1 \leq j \leq \gamma$ and $\pi_j \in \hat{e}_j$), we obtain
$$
\max_{\hat{e} \in \hat{E}_i} \sum_{\pi \in \hat{e}} \varphi(\pi)
=
\sum_{1\leq j \leq \gamma} \omega_{i,j} f(j).
$$
Hence, \eqref{eq:DAG} is true.

\item[Case 3:]  $i \in V_{\max}$.
When $i$ has only one successor relation \eqref{eq:DAG} is true because the reasoning of Case 1 applies.
So let us assume that $i$ has more than one successor. 

Let $j$ and $j'$ be two successors of $i$, and let $\hat{e} \in \hat{E}_j$ and $\hat{e}' \in \hat{E}_{j'}$. Since $i$ is a co-product there is no path of $\{(i,j)\pi : \pi \in \hat{e}  \}$ that can be compatible with a path of $\{(i,j')\pi' : \pi' \in \hat{e}'  \}$. 
Therefore, we have $\hat{e} \in \hat{E}_i$ if and only if there exists some $\hat{e}'$ such that $\hat{e}' \in \hat{E}_{j}$ and $\hat{e} = \{(i,j) \pi' : \pi' \in \hat{e}'\}$. Also, since $\omega_{i,j}=1$ we have $\sum_{\pi \in \hat{e}} \varphi(\pi)=\sum_{\pi' \in \hat{e}'} \varphi(\pi')$ and we get
$$
\max_{\hat{e} \in \hat{E}_i} \sum_{\pi \in \hat{e}} \varphi(\pi)
=
\max_{\{\hat{e}':  \hat{e}' \in \hat{E}_j, j \in \Gamma^+(i)\}} \sum_{\pi' \in \hat{e}'} \varphi(\pi'),
$$
i.e.
$$
\max_{\hat{e} \in \hat{E}_i} \sum_{\pi \in \hat{e}} \varphi(\pi)
= 
\max_{j \in \Gamma^+(i)} \max_{\hat{e}' \in \hat{E}_j} \sum_{\pi' \in \hat{e}'} \varphi(\pi').
$$
Since, we have $\mathsf{lg}(\pi') \leq k$ for $\pi' \in \hat{e}'$, the hypothesis of induction implies
$$
\max_{\hat{e} \in \hat{E}_i} \sum_{\pi \in \hat{e}} \varphi(\pi)
= 
\max_{j \in \Gamma^+(i)} f(j),
$$
and we deduce that \eqref{eq:DAG} is true.

\end{description}
~\cqfd
\begin{theo}
\label{theo:DAG}
The \MEP\ can be solved in time $\Theta(|V_s \cup V_+ \cup V_{\max} \cup V_o|+ |A|)$ when the emergy graph is a DAG, and the optimal value is obtained by
\begin{equation}
\mathsf{Em}(l,l')=\sum_{s \in V_s} f(s),
\end{equation}
where $f$ is defined by \eqref{eq:DAGfunction}.
\end{theo}
\proof
By \cref{eq:varphi1} and \cref{eq:Optim} we have $\mathsf{Em}(l,l')=\max_{\hat{\varepsilon} \in \hat{E}(l,l')} \sum_{\pi \in \hat{\varepsilon}} \varphi(\pi)$.
By definition, $\hat{E}(l,l')$ is the set of all emergy states relatively to arc $(l,l')$. 
Recalling that an emergy state is a set of pairwise compatible emergy paths, and that paths starting from different sources are always compatible (case (ii) of \cref{def:CompatiblePaths}), we have $\hat{\varepsilon} \in \hat{E}(l,l')$ if and only if there exist $\hat{\varepsilon}_s$, with $s \in V_s$, such that $\hat{\varepsilon}_s \in \hat{E}_s$ and $\hat{\varepsilon} = \cup_{s \in V_s} \hat{\varepsilon}_s$. Without  loss of generality, we assume that $V_s=\{1, \ldots,\gamma\}$ where $\gamma=|V_s|$. 

The reasoning is then almost the same as for Case 2 of \cref{lem:DAG}. Setting $E=\hat{E}_1 \times \cdots \times  \hat{E}_\gamma$, we have
%Denoting by $\{ (\hat{\varepsilon}_s) : \hat{\varepsilon}_s \in \hat{E}_s, s \in V_s \}$ the set of tuples of this bijection we get 
$$
\max_{\hat{\varepsilon} \in \hat{E}(l,l')} \sum_{\pi \in \hat{\varepsilon}} \varphi(\pi)
=
\max_{(\hat{\varepsilon}_1, \ldots, \hat{\varepsilon}_\gamma) \in E}  \sum_{\pi_j \in \cup_{j=1}^{\gamma} \hat{\varepsilon}_j} \varphi(\pi_j).
$$
Since $\hat{\varepsilon}_j \cap \hat{\varepsilon}_{j'} = \emptyset$, for $1 \leq j < j' \leq \gamma$, we get
$$
\max_{\hat{\varepsilon} \in \hat{E}(l,l')} \sum_{\pi \in \hat{\varepsilon}} \varphi(\pi)
=
\max_{(\hat{\varepsilon}_1, \ldots, \hat{\varepsilon}_\gamma) \in E} \sum_{1\leq j \leq \gamma} \ \sum_{\pi_j \in \hat{\varepsilon}_j}  \varphi(\pi_j).
$$
Applying the distributivity of operator + over $\max$ we get
$$
\max_{\hat{\varepsilon} \in \hat{E}(l,l')} \sum_{\pi \in \hat{\varepsilon}} \varphi(\pi)
=
\sum_{1\leq j \leq \gamma} \max_{\hat{\varepsilon}_j \in \hat{E}_j}  \sum_{\pi_j \in \hat{\varepsilon}_j}  \varphi(\pi_j).
$$
By \cref{lem:DAG} we have $f(j) = \max_{\hat{\varepsilon} \in \hat{E}_j}  \sum_{\pi \in \hat{\varepsilon}} \varphi(\pi)$, for $1 \leq j \leq \gamma$. Thus, $\mathsf{Em}(l,l') = \sum_{s \in V_s} f(s)$, which completes the proof.

Now, let us prove the time complexity. Let $V= V_s \cup V_+ \cup V_{\max} \cup V_o$.
First, we determine, for every $i$ of $V$,  if $\hat{\varepsilon}_i=\emptyset$ or $\hat{\varepsilon}_i \neq \emptyset$, i.e. if node $l$ is reachable from node $i$.
To do so, it suffices to compute the shortest paths from $l$ to every node $i$ in the graph $G$ where each arc is reversed. This can be done in $\Theta(|V|+|A|)$ (see, e.g., \cite[p. 655]{Cormen-Leiserson-Rivest-Stein2009}) by computing a topological order $L$ for the vertices of $V$ (i.e. if $(i_j,i_k)$ is an arc then vertex $i_j$ is before $i_k$ in $L$). 

Second, we compute $f(i)$ by considering elements $i$ of $V$ in the order of $L$: formula \eqref{eq:DAGfunction} is applied $|\Gamma^+(i)|$ times for every $i$, which leads to a total time of $\Theta(|V|+|A|)$.

Finally, computing the sum $\sum_{s \in V_s} f(s)$ takes time $\Theta(|V_s|)$, so the overall time complexity is $\Theta(|V|+ |A|)$.~\cqfd

When the emergy graph is a DAG the time-complexity of the \MEP\ does not depend on the number of emergy paths. But, in the general case, is it possible to avoid an exponential-time complexity, contrary to the algorithm proposed in \cref{sec:cograph}? We answer the question in the following section.

%%%%%%%%%%%%%%%%%%%%%%%%%%%%%%%%%%%%%%%%%%%%
%\section{Complexity of the maximum emergy flow problem}
\section{Hardness of computing the maximum empower in the general case}
\label{sec:complexity}

The \MEP\ does not seem to belong to NP since we do not know how to provide a certificate that can be checked in polynomial time. 
However, the difficulty of computing is clearly based on the set $\varepsilon(l,l')$ of emergy paths, which are simple paths when node $l'$ is not repeated. Valiant \cite{Valiant1979} proved that the problem of counting the number of simple paths between two vertices of a directed graph is a \#P-complete problem.  
We prove that Valiant's problem can be solved in polynomial time by a nondeterministic Turing machine equipped with an oracle for the \MEP.

%%%%%%%%%%%%%%%%%%%%% FIGURE #P-HARD %%%%%%%%%%
\begin{figure}[h]
\centering
\includegraphics{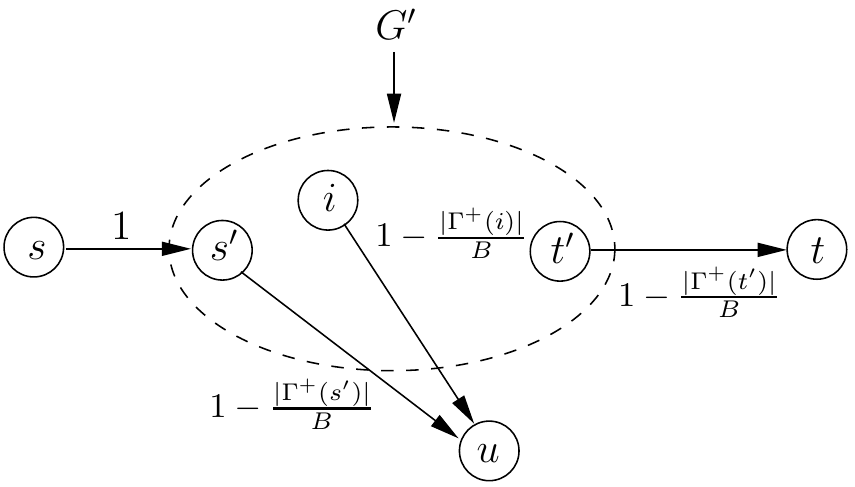}
\caption{The reduction used to prove the \#P-hardness of the \MEP.}
\label{fig:p-hard}
\end{figure}

\begin{theo}
\label{theo:P-hard}
The \MEP\ is \#P-hard.
\end{theo}

\proof 
Let $G'=(V',A',s',t')$ be an instance of the problem of counting the
number of simple paths between two vertices of a directed graph, where $s'$ and $t'$ are
the start and target vertices, respectively.
Since the number of paths
from $s'$ to $t'$ that have $i$ vertices is bounded by
$\frac{|V'|!}{(|V'|-i)!}$, the value $B=\sum_{i=1}^{|V'|}
\frac{|V'|!}{(|V'|-i)!}$ is an upper bound on the number of simple
paths from $s'$ to $t'$. We transform this instance into a \MEP\ instance as follows (see \cref{fig:p-hard}):

\begin{enumerate}
\item $V_s=\{s\}$, $V_+ = V' $, $V_{\max} = \emptyset$, and $V_o=\{ t, u \}$.

\item $A = A' \cup \{(s,s'), (t',t)\} \cup \{(i,u) : i \in V' \cup \{s'\} \}$.

\item $\omega_{i,j} = \frac{1}{B}$ for $(i, j) \in A'$.

\item $\omega_{i,u} = 1-  \frac{|\Gamma^+(i)|}{B}$ for $i \in V' \cup \{s'\}$.

\item $\omega_{s,s'} = 1$ and $\omega_{t',t} = 1-  \frac{|\Gamma^+(t')|}{B}$.

\item $\theta(s)=1$.

\end{enumerate}
Notice that we have $\sum_{j \in \Gamma^+(i)} \omega_{i,j}=1$ for $i \in V_+$, which is required for an emergy graph (recall definition of $\omega$).

We first call the oracle to get the value of $\mathsf{Em}(t',l)$. 
Since there are no co-product nodes, any set of emergy paths is an emergy state. By \eqref{eq:Optim} and because $\varphi(\pi) \ge 0$, for an emergy path $\pi$, we deduce that $\mathsf{Em}(t',t)=\sum_{\pi \in \hat{E}(t',l) } \varphi(\pi)$, where $\hat{E}(t',t)$ is the set of all emergy paths ending by arc $(t',t)$.

Then, we use this value to find the number of simple paths from $s'$ to $t'$.
Let $n_i$ be the number of elements of $\hat{E}(t',t)$ of length $i$, i.e. of emergy paths that have
exactly $i$ arcs. For such a path $\pi$ we get $\varphi(\pi)=\theta(s)\omega_{s,s'}
(\frac{1}{B})^{i-2} \omega_{t',t} = \frac{1}{B^{i-2}} \omega_{t',t}$. 
Hence, we have $\mathsf{Em}(t',t)= \sum_{i=2}^{i=|V'|+1} \frac{n_i}{B^{i-2}} \omega_{t',t} $, since $2 \leq i \leq |V'|+1$.

Now, let $E' = \frac{\mathsf{Em}(t',t)}{w_{t',t}}$.  Since $n_i < B$ (by construction), we can retrieve the value of the $n_i$'s from $E'$ by considering the representation of $E'$ in basis $\frac{1}{B}$:
by multiplying $E'$ by $B$ and taking the integer part we get $n_2$; if we substract $n_2$ to $E'$ and repeat the process we get $n_3$, and so on. Hence, we can get the value $\sum_{i=2}^{i=|V'|+1} n_i$, which is the number of simple paths from $s$ to $t$.

Since both the transformation and the computation of
$\sum_{i=2}^{i=|V'|+1} n_i$ can be done in polynomial time, we deduce
the \#P-hardness of our problem.
\cqfd

\section{Conclusion}
\label{sec:conclusion}

We have proved that the \MEP\ is \#P-hard, but solvable in polynomial-time when the emergy graph has no cycles. 
However, taking into account cycles is not only a theoretical problem because feedback arcs in the emergy graph model recycling processes, which are part of numerous real systems.
If there are cycles and the number of emergy paths is a polynomial of the size of the emergy graph, we can solve the problem in polynomial time by using the algorithm we have provided for the general case. These remarks raise the following question: 
\noi
\vskip 0.2cm
\textbf{Q1:} Is there a kind of emergy graph with cycles and an exponential number of emergy paths for which the \MEP\ can be solved in polynomial time?

A second question is related to the type of nodes in the emergy graphs, i.e. split or co-products. 
Indeed, since no co-product nodes are used in the reduction of the proof of \#P-hardness, the complexity of the problem seems to rely on the existence of split nodes (and cycles of course). 
Moreover, in case of an emergy graph with no split nodes the problem is trivially solvable: only one emergy path per source can belong to an emergy state, and the weights are all equal to 1; Hence, the problem reduces to finding a simple path from each source to the arc $(l,l')$, and the optimal value is then simply the value $\sum_{\{s: \mbox{ there is a path from } s \mbox{ to } (l,l')\}} \theta(s)$. This suggests the following problem:
\noi
\vskip 0.2cm
\textbf{Q2:} What is the complexity of the \MEP\ when the number of split nodes is fixed? 

%\section*{\refname}
%% `Elsevier LaTeX' style
\bibliographystyle{plain}
%\biboptions{}

\bibliography{ref_lt}

\begin{thebibliography}{10}

\bibitem{kn:Bac-cooq}
F.~Baccelli, G.~Cohen, G.J. Olsder, and J-P. Quadrat.
\newblock {\em {Synchronization and Linearity}}.
\newblock John Wiley and Sons, 1992.

\bibitem{kn:BackCarre75}
R.~C. Backhouse and B.~A. Carr\'e.
\newblock Regular algebra applied to path-finding problems.
\newblock {\em IMA Journ Appl. Math}, 15(2), 1975.
\newblock (161-186).

\bibitem{kn:Bardietal2005}
E.~Bardi, M.J. Cohen, and M.~T. Brown.
\newblock {A Linear Optimization Method for Computing Transformities from
  Ecosystem Energy Webs}.
\newblock {\em Emergy synthesis 3: theory and applications of the emergy
  methodologies}, 2005.
\newblock (63-74).

\bibitem{kn:Bast2011}
S.~Bastianoni, F.~Morandini, T.~Flaminio, R.~M. Pulselli, and E.~B.~P. Tiezzi.
\newblock {Emergy and Emergy Algebra Explained by Means of Ingenuous Set
  Theory}.
\newblock {\em Ecological Modelling}, 222, 2011.
\newblock (2903-2907).

\bibitem{kn:Benzaken68}
C.~Benzaken.
\newblock {Structures Alg\'ebriques des Cheminements: Pseudo-Treillis, Gerbiers
  de Carr\'e Nul}.
\newblock {\em in Network and Switching Theory}, 1968.
\newblock (40-57).

\bibitem{kn:Boltzmann1886}
L.~Boltzmann.
\newblock {Der Zweite Hauptsatz der Mechanischen W{\"a}rmetheorie}.
\newblock 1886.

\bibitem{kn:Brownetal2016}
M.~T. Brown, D.~E. Campbell, C.~De~Vilbiss, and S.~Ulgiati.
\newblock {The Geobiosphere Emergy Baseline: A Synthesis}.
\newblock {\em Ecol. Model.}, 339, 2016.
\newblock (92-95).

\bibitem{kn:Brown96}
M.~T. Brown and R.~A. Herendeen.
\newblock {Embodied Energy Analysis and Emergy Analysis: a Comparative View}.
\newblock {\em Ecological Economics}, 19, 1996.
\newblock (219-235).

\bibitem{kn:Brownetal11}
M.~T. Brown, G.~Protano, and S.~Ulgiati.
\newblock {Assessing Geobiosphere Work of Generating Global Reserves of Coal,
  Crude Oil, and Natural Gas }.
\newblock {\em Ecological Modelling}, 222(3), 2011.
\newblock (879-887).

\bibitem{kn:Campbell2016}
D.~E. Campbell.
\newblock {Emergy baseline for the Earth: A historical review of the science
  and a new calculation}.
\newblock {\em Ecol. Model.}, 339, 2016.
\newblock (96-125).

\bibitem{kn:Carre71}
B.~A. Carr\'e.
\newblock {An Algebra For Network Routing Problems}.
\newblock {\em J. Inst. Math. Appl.}, 7, 1971.
\newblock (273-294).

\bibitem{kn:Chenetal2017}
W.~Chen, W.~Liu, Y.~Geng, M.~T. Brown, C.~Gao, and R.~Wu.
\newblock {Recent Progress on Emergy Research: A Bibliometric Analysis}.
\newblock {\em Renewable and Sustainable Energy Reviews}, 73, 2017.
\newblock (1051-1060).

\bibitem{kn:Odum00}
D.~Collins and H.~T. Odum.
\newblock {Calculating Transformities With Eigenvector Method}.
\newblock {\em Emergy synthesis: theory and applications of the emergy
  methodologies}, 2000.
\newblock (265-280).

\bibitem{Cormen-Leiserson-Rivest-Stein2009}
Thomas~H. Cormen, Charles~E. Leiserson, Ronald~L. Rivest, and Clifford Stein.
\newblock {\em Introduction to algorithms}.
\newblock The MIT Press, 3rd edition, 2009.

\bibitem{Corneil-Lerchs-StewartBurlingham1981}
D.~G. Corneil, H.~Lerchs, and L.~Stewart~Burlingham.
\newblock Complement reducible graphs.
\newblock {\em Discrete Applied Mathematics}, 3:163--174, 1981.

\bibitem{kn:Vilbissetal2016}
C.~De~Vilbiss, M.~T. Brown, E.~Siegel, and S.~Arden.
\newblock {Computing The Geobiosphere Emergy Baseline: A Novel Approach}.
\newblock {\em Ecol. Model.}, 339, 2016.
\newblock (133-139).

\bibitem{kn:Giannantoni02}
C.~Giannantoni.
\newblock {\em { The Maximum Empower Principle as the Basis for Thermodynamics
  of Quality}}.
\newblock SG Editoriali, 2002.

\bibitem{kn:Giannantoni06}
C.~Giannantoni.
\newblock {Mathematics For Generative Processes: Living and Non-living
  Systems}.
\newblock {\em Journ. Comp. Appl. Math.}, 189, 2006.
\newblock (324-340).

\bibitem{Berge-Chvatal1984}
Martin~Charles Golumbic.
\newblock Topics on perfect graphs.
\newblock In Claude Berge and Vasek Chv\'atal, editors, {\em Annals of Discrete
  Mathematics}, volume~21. North-Holland, 1984.

\bibitem{kn:Horlock96}
J.H. Horlock.
\newblock {\em {Cogeneration-Combined heat and power: Thermodynamics and
  economics}}.
\newblock Krieger Publishing Company, 1996.

\bibitem{Jung1978}
H.~A. Jung.
\newblock On a class of posets and the corresponding comparability graphs.
\newblock {\em Journal of Combinatorial Theory, Series B}, 24:125--133, 1978.

\bibitem{kn:Bastia2012}
C.~Kazanci, J.~R. Schramski, and S.~Bastianoni.
\newblock {Individual Based Emergy Analysis: A Lagrangian Model of Energy
  Memory}.
\newblock {\em Ecological Complexity}, 11, 2012.
\newblock (103-108).

\bibitem{kn:Lahlou2017}
C.~Lahlou and L.~Truffet.
\newblock {Self-organization and the Maximum Empower Principle in the Framework
  of max-plus Algebra}, 2017.
\newblock arXiv:1712.05798.

\bibitem{kn:Lazz2009}
A.~Lazzaretto.
\newblock {A Critical Comparison Between Thermoeconomic and Emergy Analyses
  Algebra}.
\newblock {\em Energy}, 34, 2009.
\newblock (2196-2205).

\bibitem{kn:Lecorre2016}
O.~Le~Corre.
\newblock {\em {Emergy}}.
\newblock ISTE PRESS, 2016.

\bibitem{kn:Lecorre2012b}
O.~Le~Corre and L.~Truffet.
\newblock {A Rigourous Mathematical Framework for Computing a Sustainability
  Ratio: the Emergy}.
\newblock {\em Journal of Environmental Informatics}, 20(2), 2012.
\newblock (75-89).

\bibitem{kn:Lecorre2012}
O.~Le~Corre and L.~Truffet.
\newblock {Exact Computation of Emergy Based on a Mathematical Reinterpretation
  of the Rules of Emergy Algebra}.
\newblock {\em Ecological Modelling}, 230, 2012.
\newblock (101-113).

\bibitem{kn:Lecorre2015c}
O.~Le~Corre and L.~Truffet.
\newblock {Emergy Paths computation from Interconnected Energy System Diagram}.
\newblock {\em Ecol. Model.}, 313, 2015.
\newblock (181-200).

\bibitem{kn:Leon73}
W.~Leontief.
\newblock {\em {Input-Output economics}}.
\newblock Oxford University Press, 1973.

\bibitem{kn:Lietal2010}
L.~Li, H.~Lu, D.~E. Campbell, and H.~Ren.
\newblock {Emergy Algebra: Improving Matrix Method for Calculating
  Transformities}.
\newblock {\em Ecological Modelling}, 221, 2010.
\newblock (411-422).

\bibitem{kn:Marvugliaetal2011}
A.~Marvuglia, E.~Benetto, B.~Rugani, and G.~Rios.
\newblock {A Scalable Implementation of the Track Summing Algorithm For Emergy
  Calculation With Life Cycle Inventory Databases}.
\newblock In {\em EnviroInfo 2011}, 2011.

\bibitem{kn:Scale2013a}
A.~Marvuglia, B.~Rugani, G.~Rios, Y.~Pign{\'e}, E.~Benetto, and
  L.~Tiruta-Barna.
\newblock {Using graph search algorithms for a rigorous application of emergy
  algebra rules}.
\newblock {\em Rev. M\'etal.}, 110(1), 2013.
\newblock (87-94).

\bibitem{kn:Moranetal}
M.~J. Moran, H.~N. Shapiro, D.~D. Boettner, and M.~B. Bailey.
\newblock {\em {Fundamentals of Engineering Thermodynamics}}.
\newblock Wiley, 2014.
\newblock 8th Edition.

\bibitem{kn:Muetal12}
H.~Mu, X.~Feng, and K.~Hoong~Chu.
\newblock {Calculation of Emergy Flows Within Complex Chemical Production
  Systems}.
\newblock {\em Ecol. Engin.}, 44, 2012.
\newblock (88-93).

\bibitem{kn:Odum95}
H.~T. Odum.
\newblock {Self-organization and Maximum Empower}.
\newblock In {\em C. A. S. Hall (Ed.), Maximum Power: The Ideas and
  Applications of H. T. Odum}, 1995.
\newblock Colorado University Press, Colorado.

\bibitem{kn:Odum96}
H.~T. Odum.
\newblock {\em {Environmental accounting. EMERGY and decision making}}.
\newblock John Wiley, 1996.

\bibitem{kn:Odum55}
H.~T. Odum and R.~C. Pinkerton.
\newblock {Time's Speed Regulator: The Optimum Efficiency for Maximum Power
  Output in Physical and Biological Systems}.
\newblock {\em American Scientist}, 43(2), 1955.
\newblock (331-343).

\bibitem{kn:Patterson2014}
M.~Patterson.
\newblock {Evaluation of Matrix Algebra Methods for Calculating Transformities
  from Ecological and Economic Network Data}.
\newblock {\em Ecological Modelling}, 271, 2014.
\newblock (72-82).

\bibitem{kn:Podolinsky1880}
S.~Podolinsky.
\newblock {Le travail Humain et la Conservation de l'Energie}.
\newblock {\em Rev. Intern. des Sciences}, 5, 1880.
\newblock (57-70).

\bibitem{kn:Scienceman87}
D.~M. Scienceman.
\newblock {Energy and Emergy}.
\newblock In {\em Environmental Economics-The Analysis of a Major Interface.
  Pillet, G. and Murota, T. (eds.)}, 1987.
\newblock (257-276).

\bibitem{kn:Tennenbaum88}
S.~E. Tennenbaum.
\newblock {\em {Network energy expenditures for subsystem production}}.
\newblock PhD thesis, University of Florida, Gainesville, 1988.

\bibitem{kn:Tennenbaum2014}
S.~E. Tennenbaum.
\newblock {Emergy and Co-emergy}.
\newblock {\em Ecol. Model.}, 2014.
\newblock http://dx.doi.org/10.1016/j.ecolmodel.2014.09.012.

\bibitem{Valiant1979}
Leslie Valiant.
\newblock The complexity of enumeration and reliability problems.
\newblock {\em SIAM Journal on Computing}, 8(3):410--421, 1979.

\bibitem{kn:Valyi05}
R.~Valyi.
\newblock {About the Emergy Concept}.
\newblock Technical report, Ecole Centrale de Lyon, FRANCE, 2005.
\newblock http://emsim.sourceforge.net/latexdocs/emergy.pdf.

\bibitem{kn:Zhongetal2016}
S.~Zhong, Y.~Geng, W.~Liu, C.~Gao, and W.~Chen.
\newblock {A Bibliometric Review on Natural Resource Accounting During
  1995--2014}.
\newblock {\em Journ. Cleaner Prod.}, 139, 2016.
\newblock (122-132).

\end{thebibliography}

\end{document}